\documentclass[pdflatex,sn-mathphys-num]{sn-jnl}

\usepackage{subcaption}
\usepackage{graphicx}%
\usepackage{multirow}%
\usepackage{slashed}
\usepackage{amsmath,amssymb,amsfonts}%
\usepackage{amsthm}%
\usepackage{mathrsfs}%
\usepackage[title]{appendix}%
\usepackage{xcolor}%
\usepackage{textcomp}%
\usepackage{manyfoot}%
\usepackage{booktabs}%
\usepackage{algorithm}%
\usepackage{algorithmicx}%
\usepackage{algpseudocode}%
\usepackage{listings}%
\usepackage{longtable}

\usepackage{cancel}
\usepackage{xspace}
\newcommand{\mi}{\mathrm{i}}

\begin{document}

\title[Chiral symmetry breaking at large $N_f$ and the $\rho$ meson]{Chiral symmetry breaking for large $N_f$ and the properties of the $\rho$ meson near the conformal window}

\author*[1]{\fnm{Aftab} \sur{Ahmad}}\email{aftabahmad@gu.edu.pk}
\author[1]{\fnm{Samreen} \sur{Fatima}}\email{sf2088409@gmail.com}
\author[1]{\fnm{Muhammad} \sur{Shahbaz}}\email{m.shahbaz6544@gmail.com}
\author[2]{\fnm{Marco Antonio} \sur{Bedolla}}\email{marco.bedolla@unach.mx}
\author[3,4]{\fnm{Alfredo} \sur{Raya}}\email{alfredo.raya@umich.mx}

\author[5]{\fnm{Muhammad} \sur{Imran}}\email{ dr.Imran@ustb.edu.pk}
\affil[1]{\orgdiv{Institute of Physics}, \orgname{Gomal University},
\orgaddress{\city{D.~I.~Khan}, \postcode{29220},
\state{Khyber Pakhtunkhwa}, \country{Pakistan}}}
\affil[2]{\orgdiv{Facultad de Ciencias F\'isico-Matemáticas},
\orgname{Benem\'erita Universidad Autónoma ed Chiapas},
\orgaddress{\city{Tuxtla Guti\'errez}, \postcode{29050},
\state{Chiapas}, \country{Mexico}}}
\affil[3]{Facultad de Ingenier\'ia El\'ectrica,
Universidad Michoacana de San Nicol\'as de Hidalgo,
Edificio Omega, Ciudad Universitaria,
Morelia, Michoac\'an 58030, Mexico}
\affil[4]{Centro de Ciencias Exactas,
Universidad del B\'io-B\'io,
Casilla 447, Chill\'an, Chile}

\affil[5]{\orgdiv{Department of Physics},
\orgname{University of Science and Technology},
\orgaddress{\city{Bannu}, \postcode{29220},
\state{Khyber Pakhtunkhwa}, \country{Pakistan}}}

\abstract{We study how the ground-state properties of the $\rho$ meson respond
when the number of light-quark flavors $N_f$ is driven towards the conformal
window. The interaction is a symmetry-preserving, confining vector--vector
contact interaction whose effective coupling carries a flavor dependence
designed to encode the screening produced by fermion loops in the deep
infrared. That interaction feeds the Schwinger--Dyson equation for the
dressed-quark propagator and the homogeneous Bethe--Salpeter equation in the
rainbow-ladder truncation, regularized in the Schwinger proper-time scheme so
that quark production thresholds are absent while chiral symmetry remains
broken. Raising $N_f$ weakens the coupling, suppresses the dynamically
generated mass and restores chiral symmetry together with deconfinement at a
critical flavor number $N^{c}_{f}$. The vector channel responds differently
from the pseudoscalar one. Because the $\rho$ is not a Goldstone boson, its
mass is not protected by the symmetry that is being restored, and it simply
follows the shrinking constituent scale, while the canonically normalized
amplitude $E_\rho$ and the leptonic decay constant $f_\rho$ soften along the
way. In this framework the $\rho$ sits above the nominal $q\bar{q}$ threshold
for every flavor number examined, so the state survives only as long as
confinement removes that threshold. Dissociation is therefore governed by the
divergence of the confinement length scale rather than by a crossing of the
bound-state mass, and we contrast both criteria. The three elastic
electromagnetic form factors $G_E$, $G_M$ and $G_Q$, computed in the
generalized impulse approximation, flatten as $N_f$ grows, the zero crossing
of $G_E$ migrates towards the infrared and the charge radius expands, whereas
the magnitude of the quadrupole moment decreases. The $\rho$ thus becomes a
larger and rounder object as the theory approaches conformality.}


\maketitle

\section{Introduction}\label{section-1}

Quantum chromodynamics describes the strong interaction between quarks and
gluons, and two of its features have shaped the way the theory is used. At
short distances the coupling weakens and the quarks behave almost as free
particles, a property discovered by Gross, Wilczek and Politzer
\cite{Gross:1973id, Politzer:1973fx}. At long distances the interaction grows
instead of fading, colored objects never appear in isolation, and the spectrum
consists of hadrons \cite{Wilson:1974sk}. Between those two regimes sits
dynamical chiral symmetry breaking, the mechanism that turns a few MeV of
current mass into a constituent scale of several hundred MeV
\cite{Nambu:1961fr, Nambu:1961tp, Maskawa:1974vs, Atkinson:1986ay,
Praschifka:1987ry}. Most of the mass of visible matter originates there rather
than in the Higgs sector \cite{Ding:2022ows}, which is why the strength of the
infrared interaction, and not merely its existence, is the quantity worth
tracking.

That strength depends on how many light quark flavors circulate in the vacuum.
Gluon self-interaction produces antiscreening and quark loops produce
screening, so each additional flavor works against the mechanism that generates
mass. For a small number of flavors the antiscreening wins and the theory
confines and breaks chiral symmetry \cite{Engel:2014cka, Evans:2020ztq,
Ciambriello:2024xzd}. Above a critical value $N^{c}_{f}$ the balance tips, the
dynamically generated mass vanishes and the quarks deconfine
\cite{Appelquist:2009ty, bashir2013qcd, LSD:2014nmn}. This critical value must
lie below the flavor number at which asymptotic freedom itself is lost, which
for $SU(3)$ is $N^{AF,c}_{f}=16.5$ \cite{Politzer:1973fx}. In the interval
$N^{c}_{f}\lesssim N_f<N^{AF,c}_{f}$ an infrared fixed point governs the
dynamics and the theory becomes conformal at large distances
\cite{Caswell:1974gg, Banks:1981nn, gies2006chiral, Appelquist:2007hu,
hasenfratz2010conformal, aoki2012many, Evans:2020ztq}. Near the upper edge the
fixed point sits at weak coupling. Near the lower edge it moves into the
strongly interacting region, and the way the theory exits the conformal window
as $N_f$ decreases has been characterized as a conformal phase transition
\cite{Miransky:1996pd, Appelquist:1996dq}. Lattice simulations
\cite{LSD:2014nmn, Hayakawa:2010yn, Cheng:2013eu, Hasenfratz:2016dou,
LatticeStrongDynamics:2018hun, Deuzeman:2009mh, Aoki:2013xza, Fodor:2011tu}
and continuum methods \cite{bashir2013qcd, Appelquist:1999hr, Hopfer:2014zna,
Doff:2016jzk, Binosi:2016xxu, Ahmad:2020jzn, Ahmad:2022hbu} broadly place the
lower edge around $8\lesssim N^{c}_{f}\lesssim 12$.

Interest in this region is not confined to QCD. Gauge theories that sit just
below the conformal window run slowly over a wide range of scales, and that
walking behavior is the ingredient theories of dynamical electroweak symmetry
breaking need in order to separate the scale of fermion mass generation from
the scale of symmetry breaking \cite{Holdom:1981rm, Yamawaki:1985zg,
Appelquist:1986an, Sannino:2004qp, Dietrich:2006cm}. In such theories the
lightest vector resonance, the analogue of the $\rho$, carries much of the
phenomenological weight, since it is the state that couples to the electroweak
currents and the one collider searches would encounter first. How the mass,
the decay constant and the internal structure of a vector meson evolve as the
underlying theory approaches conformality is therefore a question with a life
beyond the 3-flavor world.

Within the continuum approach, one line of work has examined precisely this
evolution using a symmetry-preserving contact interaction whose coupling is
made to depend on $N_f$. The gap equation and the critical flavor number were
treated in Ref.~\cite{Ahmad:2020jzn}, the phase diagram at finite temperature
and density in Refs.~\cite{Ahmad:2022hbu, Ahmad:2025xfu}, where the line
separating hadronic matter from the quark gluon plasma was found to recede as
$N_f$ grows, and the Schwinger pair production rate in
Ref.~\cite{Ahmad:2023mqg}. The light pseudoscalar bound states came next
\cite{Ahmad:2024emu}. In the chiral limit the Nambu-Goldstone boson survives
until $N_f\approx 8$, and past that point its mass climbs abruptly and the
state dissolves, which is the behavior one expects from a mass that is tied to
the order parameter of the symmetry being restored. The vector channel has no
such protection, and what happens there cannot be inferred from the
pseudoscalar result. That gap is what the present work addresses.

The $\rho$ meson is a natural place to start. It carries $J^{PC}=1^{--}$, a
mass of $775.26\pm 0.23$ MeV and a width near $147$ MeV
\cite{ParticleDataGroup:2024cfk}, it shares its valence content with the pion
with the quark spins aligned rather than antialigned, and it decays almost
entirely into a charged pion pair. On the experimental side, diffractive
$\rho^{0}$ electroproduction was measured at HERA \cite{ZEUS:2007ipf},
exclusive photoproduction in ultraperipheral heavy-ion collisions has been
reported by ALICE \cite{ALICE:2015nbw}, and the Electron-Ion Collider is
designed to map the same channel with far greater precision
\cite{AbdulKhalek:2021gbh}. On the theoretical side a spin-1 target is more
demanding than a spin-0 one, because elastic electromagnetic scattering
requires 3 form factors instead of a single one, and the quadrupole form
factor carries information about the deformation of the charge distribution
that has no pseudoscalar analogue \cite{Bhagwat:2006pu, Hedditch:2007ex,
Owen:2015gva, Carrillo-Serrano:2015uca}.

We compute the mass, the leptonic decay constant, the canonically normalized
Bethe-Salpeter amplitude and the 3 Sachs form factors of the $\rho$ as
functions of $N_f$, together with the charge radius, the magnetic moment and
the quadrupole moment that follow from them. The dressed-quark propagator comes
from the Schwinger-Dyson equation and the bound state from the homogeneous
Bethe-Salpeter equation \cite{Salpeter:1951sz}, which in a given $J^{PC}$
channel is an eigenvalue problem whose solution satisfies
$\lambda(P^2=-m^{2}_{H})=1$, with $P$ the total momentum and $m_H$ the mass of
the state \cite{Bedolla:2015mpa, Maris:1997tm}. The eigenvector is the
Bethe-Salpeter amplitude, the object that governs production and scattering
processes involving the meson. Both equations are built on the confining
vector-vector contact interaction of Refs.~\cite{GutierrezGuerrero:2010md,
Roberts:2011cf, Roberts:2011wy}, with the flavor dependence introduced in
Refs.~\cite{Ahmad:2020jzn, Ahmad:2022hbu, Ahmad:2023mqg}, working in Landau
gauge within the rainbow-ladder truncation and regularizing with the Schwinger
proper-time scheme, which preserves the relevant Ward-Takahashi identities and
leaves a sensible constituent-like mass in the infrared.

One feature of the vector channel forces a change in how dissociation is
diagnosed, and it is worth stating early. In the contact interaction the $\rho$
lies above the sum of the constituent masses already at the physical flavor
number, a well known consequence of the momentum-independent kernel, and it
remains bound only because the infrared cutoff of the proper-time scheme
removes quark production thresholds altogether. A criterion based on the
bound-state mass climbing to meet $2M_f$, which is the natural one in the
pseudoscalar channel, therefore never triggers here. What does happen is that
the confinement length scale diverges when the dynamically generated mass
vanishes, the thresholds reappear, and the excess that was always present turns
into an open decay channel. We report both criteria and contrast them, since
the comparison clarifies what the model is and is not able to say about the
fate of a vector state near conformality.
A caveat about the meaning of the flavor number conditions everything that
follows. Varying $N_f$ here does not alter the valence content of the meson.
The situation parallels quenched QCD, where setting $N_f=0$ removes the
vacuum-polarization insertions, which are proportional to $N_f$, without
removing quarks or hadrons from the spectrum. In our case $N_f$ enters the
Feynman rules as a parameter while the quark propagators and the bound-state
equations are treated consistently, so raising it above its physical value
increases the fermion-loop contribution and the associated screening, leaving
the internal structure and the quantum numbers of the state untouched. What the
calculation probes is screening dynamics in a family of QCD-like theories,
rather than the flavor composition of physical hadrons \cite{Ahmad:2024emu}.
To this end, we have organized the remaining as follows: Section~\ref{section-2} 
sets out the flavor-dependent contact interaction and
the gap equation. Section~\ref{section-3} develops the Bethe-Salpeter equation
for the vector channel, the dressed quark-photon vertex, the canonical
normalization and the form factors. Section~\ref{section-4} presents the
numerical results and Section~\ref{section-5} closes with the perspectives they
open.

\section{Flavor-dependent contact interaction and the gap equation}\label{section-2}

\begin{figure}
\begin{center}
\includegraphics[width=9cm]{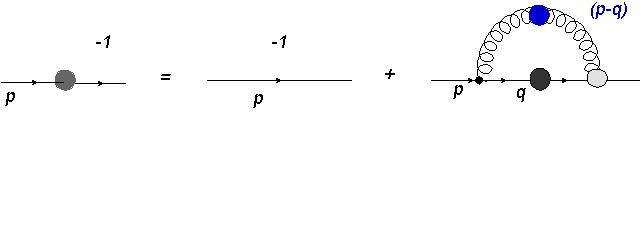}
\caption{Schwinger-Dyson equation for the dressed quark, Eq.~(\ref{CI1}). The
left-hand side is the inverse dressed propagator carrying momentum $p$. On the
right, the first term is the inverse bare propagator and the second is the
self-energy, built from the dressed gluon propagator with momentum $p-q$ (gluon
line with a blue blob), the dressed quark-gluon vertex (gray blob) and the
dressed quark propagator with momentum $q$ (line with a solid black blob).}
\label{SDEf}
\end{center}
\end{figure}

The dressed-quark propagator obeys the Schwinger-Dyson equation
\cite{Schwinger:1951nm}, depicted in Fig.~\ref{SDEf},
\begin{eqnarray}
S^{-1}_{f}(p) = S^{-1}_{f,0}(p) + \Sigma(p)\,,\label{CI1}
\end{eqnarray}
where $S_{f,0}(p)=(\slashed{p}+m_f+\mi\epsilon)^{-1}$ is the bare propagator in
Euclidean space with signature $(+,+,+,+)$, $S_{f}(p)=(\slashed{p}+M_f+\mi
\epsilon)^{-1}$ is the dressed one, $m_f$ is the current mass and $M_f$ the
dressed mass. The self-energy reads
\begin{eqnarray}
\Sigma(p)=\int\frac{d^4q}{(2\pi)^4}\,g^{2}D_{\mu\nu}(p-q)
\frac{\lambda^a}{2}\gamma_\mu S_{f}(q)\frac{\lambda^a}{2}\Gamma_\nu(p,q)\,,
\label{CI2}
\end{eqnarray}
with $g^2$ the running coupling attached to the quark-gluon vertex,
$\Gamma_\nu$ the dressed vertex, $D_{\mu\nu}$ the gluon propagator and $p-q$
the gluon momentum. The Dirac matrices satisfy
$\{\gamma_\mu,\gamma_\nu\}=2\delta_{\mu\nu}$ and
$\gamma_\mu^{\dagger}=\gamma_\mu$, and Euclidean scalar products are
$a\cdot b=\sum_{i=1}^{4}a_ib_i$. The color matrices $\lambda^a$, with
$a=1,\dots,8$, obey
\begin{eqnarray}
\sum^{N_{c}^{2}-1}_{a=1}\frac{{\lambda}^a}{2}\frac{{\lambda}^a}{2}
=\frac{1}{2}\left(N_c-\frac{1}{N_c}\right)I\,,\label{CI3}
\end{eqnarray}
with $I$ the identity in color space and $N_c$ the number of colors.

\begin{figure}
\begin{center}
\includegraphics[width=6cm]{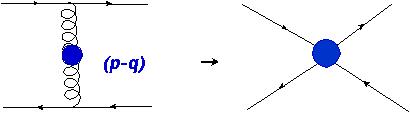}
\caption{The contact interaction of Eq.~(\ref{CI4}). A constant gluon
propagator turns one-gluon exchange into an effective four-quark coupling.}
\label{KCI}
\end{center}
\end{figure}

Solving Eq.~(\ref{CI1}) requires a truncation. We adopt the symmetry-preserving
contact interaction \cite{GutierrezGuerrero:2010md, Roberts:2011cf,
Roberts:2011wy, Qin:2011dd, Chen:2012qr}, in which the vertex reduces to
$\Gamma_\nu=\gamma_\nu$ and the Landau-gauge gluon propagator is evaluated in
the deep infrared, where lattice and continuum studies agree that the gluon
acquires a dynamically generated mass $m_g$ \cite{Langfeld:1996rn,
Cornwall:1981zr, PhysRevD.80.085018, Aguilar:2015bud, Oliveira:2010xc,
Boucaud:2010gr, Kohyama:2016obc, Ferreira:2025anh}. Both the propagator and the
coupling are then treated as momentum independent \cite{GutierrezGuerrero:2010md,
Roberts:2011cf, Roberts:2011wy, Chen:2012qr, Xing:2021dwe, Deur_2024},
\begin{eqnarray}
g^2 D_{\mu\nu}(p-q)\rightarrow\frac{4\pi\alpha_{\rm ir}}{m_{g}^{2}}\,
\delta_{\mu\nu}\doteq\delta_{\mu\nu}\,\alpha_{\rm eff}\,.\label{CI4}
\end{eqnarray}
The one-gluon exchange collapses into the effective four-quark coupling of
Fig.~\ref{KCI}. Despite its simplicity, this interaction reproduces the static
properties of light pseudoscalar and vector mesons at a level comparable with
far more elaborate kernels \cite{Maris:2006ea, Bashir:2012fs, Eichmann:2008ae,
Cloet:2007pi}.

The infrared strength $\alpha_{\rm ir}=0.93\pi$ is consistent with current
estimates of the zero-momentum value of the QCD running coupling
\cite{Maris:2005tt, Aguilar:2010gm, Brodsky:2010px, Deur:2016tte,
Aguilar:2015bud, Deur_2024}, and the gluon mass scale $m_g=0.8$ GeV follows the
one-loop renormalization-group-improved interaction of
Refs.~\cite{Qin:2011dd, Chen:2012qr, Bedolla:2015mpa}. Both were fixed long ago
by light-hadron phenomenology \cite{GutierrezGuerrero:2010md, Roberts:2011cf,
Roberts:2011wy}. The normalization used here is related to the one employed in
that earlier literature by $4\pi\alpha_{\rm ir}/m^{2}_{g}=1/m^{2}_{G}$ with
$m_G=0.132$ GeV \cite{Wang:2013wk, Bedolla:2015mpa}. With this kernel the mass
function loses its momentum dependence and becomes a constant.

\subsection{Flavor dependence of the effective coupling}

The contact interaction as it stands knows nothing about how many quark flavors
populate the vacuum. To build that information in, Refs.~\cite{Ahmad:2020jzn,
Ahmad:2022hbu, Ahmad:2025xfu} modified $\alpha_{\rm eff}$ so that the gap
equation admits solutions over a range of flavor numbers and the dynamically
generated mass approaches its critical point in the manner expected from
Schwinger-Dyson studies with two different truncations \cite{bashir2013qcd},
\begin{eqnarray}
M_f\sim\sqrt{1-\frac{N_f}{N^{c}_{f}}}\,.\label{CI4a}
\end{eqnarray}
Reproducing that square-root approach inside a four-fermion interaction
requires a square-root dependence in the coupling itself
\cite{Ahmad:2020jzn, Ahmad:2022hbu, Ahmad:2025xfu},

\begin{eqnarray}
\alpha_{\rm eff}(N_f)=\alpha_{\rm eff}
\sqrt{1-\frac{(N_{f}-2)}{\mathcal{N}_{f}^{c}}}\,.\label{CI4b}
\end{eqnarray}
At the physical flavor number $N_f=2$ the flavor-dressed coupling collapses onto
the original $\alpha_{\rm eff}$, so nothing that was fitted to light-hadron data
is disturbed. The quantity $\mathcal{N}^{c}_{f}=N^{c}_{f}+\varsigma$ is a trial
critical number whose offset $\varsigma$ absorbs the shift produced by the
factor $N_f-2$ \cite{Ahmad:2020jzn}. Setting $\varsigma=2.9$ places the critical
flavor number in the chiral limit at $N^{c}_{f}\big|_{\rm chiral}\approx 8$,
above which dynamical chiral symmetry is restored and the quarks deconfine. With
the current masses of Table~\ref{tab1} the restoration is only partial and the
corresponding value shifts to $N^{c}_{f}\big|_{m_f\neq 0}\approx 8.2$. We keep
that distinction explicit throughout, since the 2 numbers refer to different
physical situations.
The subtraction of $2$ in Eq.~(\ref{CI4b}) is a bookkeeping device tied to
$\varsigma$, not a statement about physical flavor content, because what
multiplies the vacuum polarization is the number of active loops rather than the
number of quarks in the meson. Nothing prevents one from writing $N_f-1$
instead, at the price of retuning $\varsigma$ to land on the same critical
number. The choice matters little for the strange quark in particular, since the
parameters were fitted to light-hadron observables built from $u\bar{d}$,
$u\bar{s}$ and similar combinations, rather than to any single quark mass. When
chiral symmetry breaking and the conformal window are at issue, only the up,
down and strange flavors remain active in the infrared, while charm, bottom and
top are effectively frozen out and can be dropped.

What Eq.~(\ref{CI4b}) encodes physically is a competition. For few light
flavors, gluon self-interaction produces an antiscreening effect that outweighs
the screening from fermion loops, the theory is strongly coupled in the deep
infrared, and it responds by generating mass and confining. Adding flavors adds
quark loops, screening gains ground, the effective interaction weakens, and
beyond a point it can no longer sustain either mass generation or confinement.
The same trend survives across the several variants of this construction that
have appeared in the literature, so the persistence or eventual loss of
dynamical chiral symmetry as $N_f$ moves into the interval
$N^{c}_{f}\lesssim N_f<N^{AF,c}_{f}$ is a robust feature rather than an artifact
of one particular parametrization. Its consequences for the bound-state spectrum
are what we set out to quantify.
Perturbation theory points the other way, and the contrast is instructive. At a
fixed energy scale the perturbative running coupling grows with $N_f$ once the
flavor dependence of the $\beta$-function coefficients is taken into account. The
sign of the one-loop coefficient is controlled by $11N_c-2N_f$, which for
$N_c=3$ and a handful of flavors is negative, hence asymptotic freedom
\cite{Gross:1973id, Politzer:1973fx}. Past $N_f=16.5$ in $SU(3)$ the sign flips,
the coupling grows in the ultraviolet and dies in the infrared, screening
dominates at every scale, and the theory is presumably trivial. At intermediate
flavor numbers antiscreening still governs the far ultraviolet, and as the scale
is lowered the coupling grows but quark-loop screening keeps it from reaching the
critical strength needed to break chiral symmetry \cite{Hopfer:2014zna}. No mass
is generated and the light, long-ranged quarks continue to dominate the
nonperturbative dynamics. The flavor-dependent coupling of Eq.~(\ref{CI4b}) is
not meant to reproduce that perturbative running. It is a modelling device for
the infrared, tuned so that the dynamical mass follows Eq.~(\ref{CI4a}), and it
should be read as such.

\subsection{The gap equation}

Replacing Eq.~(\ref{CI4b}) in Eq.~(\ref{CI4}) gives the flavor-dependent contact
interaction,
\begin{eqnarray}
g^2 D_{\mu\nu}(p-q)\rightarrow\delta_{\mu\nu}\,\alpha_{\rm eff}(N_f)\,,
\label{CI4c}
\end{eqnarray}
and with $N_c=3$, after tracing over the Dirac indices, the dressed mass obeys
\begin{equation}
M_f=m_f+\frac{4\alpha_{\rm eff}(N_f)}{3}\int\frac{d^{4}q}{(2\pi)^4}\,
{\rm Tr}\left[S_{f}(q)\right]\,,\label{CI6}
\end{equation}
that is,
\begin{eqnarray}
M_f=m_f+\frac{16\alpha_{\rm eff}(N_f)}{3}\int\frac{d^4q}{(2\pi)^4}
\frac{M_f}{q^2+M^{2}_{f}}\,.\label{CI7}
\end{eqnarray}
The structure is that of a mean-field equation of the Nambu-Jona-Lasinio type
\cite{Nambu:1961fr}, in which the four-fermion interaction is traded for a
constant condensate that supplies the quark with a self-consistently determined
mass. Writing $s=q^2$ and
$d^4q=(1/2)q^2dq^2\sin^2\theta\,d\theta\sin\phi\,d\phi\,d\psi$, the angular
integrals give
\begin{eqnarray}
M_f=m_{f}+\frac{M_f\,\alpha_{\rm eff}(N_f)}{3\pi^2}\int^{\infty}_{0}
ds\,\frac{s}{s+M_f^2}\,.\label{CI9}
\end{eqnarray}

This integral diverges and the model must be regularized. We use the Schwinger
proper-time scheme \cite{Schwinger:1951nm}, which exponentiates the denominator
and cuts the proper-time interval at both ends, with an ultraviolet cutoff
$\tau_{uv}=1/\Lambda_{uv}$ and an infrared one $\tau_{ir}=1/\Lambda_{ir}$,
\begin{eqnarray}
\frac{1}{s+M^{2}_{f}}&=&\int^{\infty}_{0}d\tau\,
{\rm e}^{-\tau(s+M^{2}_{f})}\rightarrow
\int^{\tau_{ir}^2}_{\tau_{uv}^2}d\tau\,{\rm e}^{-\tau(s+M^{2}_{f})}
\nonumber\\
&=&\frac{{\rm e}^{-\tau_{uv}^2(s+M^{2}_{f})}
-{\rm e}^{-\tau_{ir}^2(s+M^{2}_{f})}}{s+M^{2}_{f}}\,.\label{CI10}
\end{eqnarray}
The infrared cutoff is what implements confinement, since it removes the quark
production thresholds that an unregulated propagator would exhibit
\cite{Ebert:1996vx, Roberts:2007jh, Roberts:2011cf, Roberts:2011wy}. This point
carries most of the weight in the vector channel and we return to it in
Sec.~\ref{section-4}. The ultraviolet cutoff plays a different role. The contact
interaction is not renormalizable, so $\tau_{uv}$ cannot be removed and becomes
part of the model, setting the scale of every dimensionful quantity in it.
Raising $\Lambda_{uv}$ is also how one mimics the short-distance effects that
matter for heavier quarks. The scheme eliminates quadratic and logarithmic
divergences and respects the axial-vector Ward-Takahashi identity
\cite{Ward:1950xp, Takahashi:1957xn}, which is what makes it usable for
bound-state calculations. Performing the integral over $s$ leaves
\begin{eqnarray}
M_f=m_{f}+\frac{M_f\,\alpha_{\rm eff}(N_f)}{3\pi^{2}}\,
\mathcal{A}_{01}\!\left(M_f^2;\tau_{\rm uv}^2,\tau_{\rm ir}^2\right)
\,,\label{CI11}
\end{eqnarray}
where the family of regularized integrals is
\begin{equation}
\mathcal{A}_{\delta\zeta}\!\left(M_f^2;\tau_{\rm uv}^{2},\tau_{\rm ir}^{2}\right)
=\frac{(M^{2}_{f})^{\epsilon}}{\Gamma(\zeta)}\,
\Gamma\!\left(\zeta-2,\tau_{\rm uv}^{2}M_f^2,\tau_{\rm ir}^{2}M_f^2\right)
\,,\label{CI12}
\end{equation}
with $\epsilon=\delta-(\zeta-2)$,
$\Gamma(a,y_1,y_2)=\Gamma(a,y_1)-\Gamma(a,y_2)$ and
$\Gamma(a,y)=\int_{y}^{\infty}t^{a-1}{\rm e}^{-t}dt$ the incomplete Gamma
function. The combination that appears in Eq.~(\ref{CI11}) is the one with
$\delta=0$ and $\zeta=1$, for which $\epsilon=1$ and
$\mathcal{A}_{01}=M_f^{2}\left[\Gamma(-1,\tau_{\rm uv}^{2}M_f^{2})
-\Gamma(-1,\tau_{\rm ir}^{2}M_f^{2})\right]$.

Confinement in this model is monitored through a flavor-dependent length scale
\cite{Ahmad:2016iez, Ahmad:2020ifp, Ahmad:2020jzn, Ahmad:2023mqg,
Ahmad:2023ecw, Ahmad:2025prr, Ahmad:2025xfu},
\begin{eqnarray}
\tilde{\tau}_{ir}=\tau_{ir}\,\frac{M(2)}{M(N_f)}\,,\label{CI13}
\end{eqnarray}
where $M(2)$ is the dressed mass at $N_f=2$ and $M(N_f)$ its generalization. The
regulator that ties chiral symmetry breaking to confinement therefore inherits an
explicit dependence on the flavor number. In the chiral limit $M(N_f)$ vanishes
at $N^{c}_{f}$ and $\tilde{\tau}_{ir}$ diverges there, the production thresholds
reopen, and the system deconfines \cite{Ahmad:2020jzn, Ahmad:2023mqg,
Ahmad:2025xfu}. That divergence is the quantity we will use to decide when the
vector bound state ceases to exist.

\section{Bethe-Salpeter equation for vector mesons}\label{section-3}

\begin{figure}
\begin{center}
\includegraphics[width=9cm]{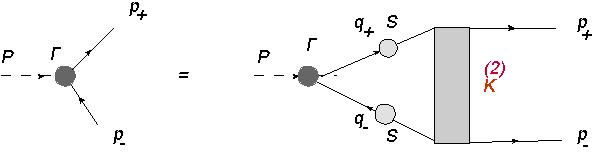}
\caption{The Bethe-Salpeter equation, Eq.~(\ref{BSE1}). Light gray blobs are
dressed quark propagators $S_f$, black solid circles are the Bethe-Salpeter
amplitude $\Gamma_{H}$, and the gray rectangle is the dressed quark-antiquark
kernel $K^{(2)}$.}
\label{BSEfig}
\end{center}
\end{figure}

Bound states appear as poles of the four-point quark-antiquark function, and the
homogeneous Bethe-Salpeter equation fixes the mass and the internal structure of
the state in a given $J^{PC}$ channel. For a vector meson $H=q\bar{q}$ with total
momentum $P$ and relative momentum $p$, that equation reads
\cite{Salpeter:1951sz, Maris:1997hd}
\begin{equation}
\Gamma_{H\mu}(p;P)=\int\frac{d^{4}q}{(2\pi)^4}\,K^{(2)}(p,q;P)\,
\chi_{H\mu}(q;P)\,,\label{BSE1}
\end{equation}
with the Bethe-Salpeter wave function
\begin{equation}
\chi_{H\mu}(q;P)=S_{f}(q_{+})\Gamma_{H\mu}(q;P)S_{g}(q_{-})\,.\label{BSE1a}
\end{equation}
Here $q_{+}=q+uP$ and $q_{-}=q-(1-u)P$, with $u\in[0,1]$ the momentum-sharing
parameter, which we set to $u=1/2$ for the equal-mass system at hand. Physical
observables do not depend on that choice. The propagators $S_{f,g}$ carry the two
quark flavors and $K^{(2)}(p,q;P)$ is the two-body irreducible quark-antiquark
scattering kernel, sketched in Fig.~\ref{BSEfig}.

\begin{figure}
\begin{center}
\includegraphics[width=9cm]{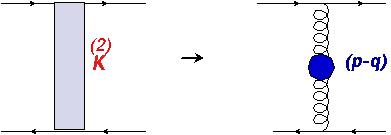}
\caption{The scattering kernel $K^{(2)}$ of Eq.~(\ref{BSE3a}) in ladder
approximation.}
\label{KBSE}
\end{center}
\end{figure}

Vector and axial-vector Ward-Takahashi identities are preserved only if
$K^{(2)}$ is built consistently with the self-energy kernel of the gap equation.
In rainbow-ladder truncation with the interaction of Eq.~(\ref{CI4c}), the
two-body kernel inherits the same momentum independence,
\begin{equation}
K^{(2)}(p,q;P)=-\frac{4}{3}\alpha_{\rm eff}(N_f)\,\delta_{\alpha\beta}\,
\gamma_\alpha\otimes\gamma_\beta=-K^{(1)}(q,p;P)\,,\label{BSE3a}
\end{equation}
where $K^{(1)}$ is the rainbow piece attached to the propagator and $K^{(2)}$ the
ladder piece attached to the amplitude \cite{Xing:2021dwe}, drawn in
Fig.~\ref{KBSE}. A kernel that does not depend on the relative momentum produces
an amplitude $\Gamma_{H\mu}$ that does not either, which is the pointlike
composite structure characteristic of contact interactions at lowest order
\cite{GutierrezGuerrero:2010md, Roberts:2011cf}.

\subsection{Amplitude of the \mbox{\boldmath $\rho$} meson}

For $J^{PC}=1^{--}$ the amplitude decomposes into 2 transverse Dirac
structures 
\cite{Llewellyn-Smith:1969bcu, Maris:1999nt}
\begin{equation}
\Gamma_{\mu}^\rho(P)=\gamma^{T}_{\mu}E_\rho(P)
+\frac{1}{2M_{f}}\sigma_{\mu\nu}P_{\nu}F_\rho(P)\,,\label{rhobsa}
\end{equation}
with $\gamma^T_\mu=\gamma_\mu-P_\mu\slashed{P}/P^2$, which satisfies
$P_\mu\gamma^T_\mu=0$, and
$\sigma_{\mu\nu}=(\mi/2)[\gamma_\mu,\gamma_\nu]$. The scale in the tensor
covariant is the dressed quark mass. Its precise choice is immaterial here,
because a symmetry-preserving contact interaction in rainbow-ladder truncation
forces the tensor amplitude to vanish identically \cite{Roberts:2011cf,
Bedolla:2015mpa},
\begin{equation}
F_\rho(P)\stackrel{\mbox{\footnotesize RL}}{\equiv}0\,.\label{Frhozero}
\end{equation}
Beyond rainbow-ladder, or with a momentum-dependent gluon propagator,
$F_\rho$ is nonzero and the prefactor would have to be settled. Within our
truncation the single scalar function $E_\rho(P)$ carries all the information
about the state, and inserting Eq.~(\ref{rhobsa}) into Eq.~(\ref{BSE1}) turns the
bound-state problem into an eigenvalue condition that, as we show next,
coincides with the pole of the transverse part of the dressed quark-photon
vertex.

\subsection{Dressed quark-photon vertex}

Electromagnetic form factors require a quark-photon vertex $\Gamma_\mu(Q)$
consistent with the vector Ward-Takahashi identity,
\begin{equation}
Q_\mu\Gamma_\mu(Q)=S^{-1}(k_+)-S^{-1}(k_-)\,,\label{VWTI}
\end{equation}
with $Q=k_+-k_-$ the incoming photon momentum. Gauge invariance together with
rainbow-ladder consistency demands that the vertex solve the inhomogeneous
Bethe-Salpeter equation
\begin{equation}
\Gamma_\mu(Q)=\gamma_\mu-\frac{4}{3}\alpha_{\rm eff}(N_f)
\int\frac{d^4q}{(2\pi)^4}\gamma_\alpha S(q+Q)\Gamma_\mu(Q)S(q)\gamma_\alpha\,.
\label{GammaQeq}
\end{equation}
The momentum-independent kernel splits the vertex into longitudinal and
transverse pieces,
\begin{equation}
\Gamma_\mu(Q)=\gamma^T_\mu P_T(Q^2)+\gamma_\mu^L P_L(Q^2)\,,\label{GammaQ}
\end{equation}
with $\gamma_\mu^L=Q_\mu\slashed{Q}/Q^2$ and
$\gamma^T_\mu=\gamma_\mu-\gamma_\mu^L$. Contracting Eq.~(\ref{GammaQeq}) with
$Q_\mu$ and using the identities above gives
\begin{equation}
P_L(Q^2)=1\,,\label{PL0}
\end{equation}
which keeps the photon massless and charge conserved. The transverse dressing
follows from the same equation,
\begin{equation}
P_T(Q^2)=\frac{1}{1+K_\gamma(Q^2)}\,,\label{PTQ2}
\end{equation}
with
\begin{equation}
K_\gamma(Q^2)=\frac{\alpha_{\rm eff}(N_f)}{3\pi^2}\int_0^1 d\alpha\,
\alpha(1-\alpha)\,Q^2\,
\overline{\mathcal{C}}^{\rm iu}_1\!\left(\omega(M_f^2,\alpha,Q^2)\right)\,,
\label{Kgamma}
\end{equation}
where the Feynman-parametrized scale is
\begin{equation}
\omega(M_f^2,\alpha,Q^2)=M_f^2+\alpha(1-\alpha)Q^2\,,\label{eq:omega}
\end{equation}
and the regularized proper-time functions are
\begin{eqnarray}
\overline{\mathcal{C}}^{\rm iu}_1(z)&=&\frac{\mathcal{C}^{\rm iu}_1(z)}{z}
=\Gamma(0,z\tau_{\rm uv}^2)-\Gamma(0,z\tau_{\rm ir}^2)\,,\label{eq:C1bar}\\
\mathcal{C}^{\rm iu}_1(z)&=&-z\frac{d}{dz}\mathcal{C}^{\rm iu}(z)
=z\left[\Gamma(0,z\tau_{\rm uv}^2)-\Gamma(0,z\tau_{\rm ir}^2)\right]\,.
\label{eq:C1}
\end{eqnarray}
For spacelike momenta the argument of these functions is positive and everything
is manifestly real. For timelike momenta $\omega$ can turn negative, and each
incomplete Gamma function acquires an imaginary part, but the two contributions
cancel in the difference of Eq.~(\ref{eq:C1bar}). The absence of an imaginary
part is precisely the statement that the regularization has removed the
production threshold, which is how confinement operates in this scheme and why
the vector channel can be followed to timelike momenta at all.

Since $K_\gamma(0)=0$, Eq.~(\ref{PTQ2}) gives $P_T(0)=1$ and the dressed vertex
reduces to the bare one at zero momentum transfer, as gauge invariance requires.
Moving to timelike momenta, $P_T$ develops a pole at $Q^2=-m_\rho^2$. The
ground-state mass of the $\rho$ is therefore the solution of
\begin{equation}
1+K_\gamma(-m_\rho^2)=0\,.\label{rhobse}
\end{equation}

\subsection{Normalization and leptonic decay constant}

The amplitude must be canonically normalized so that the bound-state pole in the
four-point function has unit residue. In rainbow-ladder truncation with the
contact interaction that condition reduces to
\begin{equation}
\frac{1}{E_\rho^2}=\frac{9}{\alpha_{\rm eff}(N_f)}
\left.\frac{d}{dQ^2}K_\gamma(Q^2)\right|_{Q^2=-m_\rho^2}\,,\label{Erhonorm}
\end{equation}
where the derivative is taken with respect to the photon virtuality and
evaluated at the pole.

The leptonic decay constant measures the coupling of the vector current to the
meson and is defined by the matrix element
\begin{equation}
f_\rho m_\rho\,\delta^{T}_{\mu\nu}(P)=\langle 0|\bar{q}\gamma_\mu q|
\rho_\nu(P)\rangle=N_c\int\frac{d^4q}{(2\pi)^4}\,
{\rm Tr}\left[\gamma_\mu S(q_+)\Gamma_\nu^\rho(P)S(q_-)\right]\,,
\label{frhodef}
\end{equation}
with $\delta^{T}_{\mu\nu}(P)=\delta_{\mu\nu}-P_\mu P_\nu/P^2$ the transverse
projector. Carrying out the Dirac trace, introducing a Feynman parameter and
regularizing as in Eq.~(\ref{CI10}) leaves
\begin{equation}
f_\rho=\frac{N_c\,E_\rho}{4\pi^2 m_\rho}\int_0^1 d\alpha
\left[\mathcal{A}_{01}(\omega_1)
+\left(M_f^2+\alpha(1-\alpha)m_\rho^2\right)
\overline{\mathcal{C}}^{\rm iu}_1(\omega_1)\right]\,,\label{frhoint}
\end{equation}
with $\omega_1=\omega(M_f^2,\alpha,-m_\rho^2)=M_f^2-\alpha(1-\alpha)m_\rho^2$.
The first term inside the bracket is the quadratically divergent piece of the
loop, which the vector Ward-Takahashi identity removes from any transverse
quantity. Once that cancellation is enforced, the remaining integral is the
residue of the transverse dressing function at its pole, and the decay constant
can be written in closed form as
\begin{equation}
f_\rho=-\frac{9}{2}\,\frac{E_\rho}{m_\rho\,\alpha_{\rm eff}(N_f)}\,
K_\gamma(-m_\rho^2)=\frac{9}{2}\,\frac{E_\rho}{m_\rho\,\alpha_{\rm eff}(N_f)}\,,
\label{fpim}
\end{equation}
where the last step uses $K_\gamma(-m_\rho^2)=-1$ from Eq.~(\ref{rhobse}). This
is the expression used in the numerical work of Sec.~\ref{section-4}.

\subsection{Elastic electromagnetic form factors}

A spin-1 meson has 3 elastic electromagnetic form factors, electric
$G_E(Q^2)$, magnetic $G_M(Q^2)$ and quadrupole $G_Q(Q^2)$
\cite{Bhagwat:2006pu}. With $Q$ the incoming photon momentum and
$p^i=K-Q/2$, $p^f=K+Q/2$ the initial and final meson momenta, subject to
$K\cdot Q=0$ and $K^2+Q^2/4=-m_\rho^2$, the $\rho$-$\gamma$-$\rho$ vertex
carries 3 independent tensor structures,
\begin{equation}
\Lambda_{\lambda,\mu\nu}(K,Q)=\sum_{j=1}^{3}T_{\lambda,\mu\nu}^j(K,Q)\,
F_j(Q^2)\,,\label{Lambdarho}
\end{equation}
namely
\begin{eqnarray}
T_{\lambda,\mu\nu}^1(K,Q)&=&2K_\lambda\,\mathcal{P}^T_{\mu\alpha}(p^i)\,
\mathcal{P}^T_{\alpha\nu}(p^f)\,,\\
T_{\lambda,\mu\nu}^2(K,Q)&=&\left[Q_\mu-p^i_\mu\frac{Q^2}{2m_\rho^2}\right]
\mathcal{P}^T_{\lambda\nu}(p^f)
-\left[Q_\nu+p^f_\nu\frac{Q^2}{2m_\rho^2}\right]
\mathcal{P}^T_{\lambda\mu}(p^i)\,,\\
T_{\lambda,\mu\nu}^3(K,Q)&=&\frac{K_\lambda}{m_\rho^2}
\left[Q_\mu-p^i_\mu\frac{Q^2}{2m_\rho^2}\right]
\left[Q_\nu+p^f_\nu\frac{Q^2}{2m_\rho^2}\right]\,,
\end{eqnarray}
with $\mathcal{P}^T_{\mu\nu}(p)=\delta_{\mu\nu}-p_\mu p_\nu/p^2$. The
Ward-Takahashi identities impose $Q_\lambda\Lambda_{\lambda,\mu\nu}=0$ and
$p^i_\mu\Lambda_{\lambda,\mu\nu}=0=p^f_\nu\Lambda_{\lambda,\mu\nu}$, which is a
useful check on the numerical evaluation.

The Sachs form factors follow from the $F_j$ through
\begin{eqnarray}
G_E(Q^2)&=&F_1(Q^2)+\frac{2}{3}\eta\,G_Q(Q^2)\,,\label{GEdef}\\
G_M(Q^2)&=&-F_2(Q^2)\,,\label{GMdef}\\
G_Q(Q^2)&=&F_1(Q^2)+F_2(Q^2)+\left[1+\eta\right]F_3(Q^2)\,,\label{GQdef}
\end{eqnarray}
with $\eta=Q^2/(4m_\rho^2)$. At vanishing momentum transfer they reduce to the
static properties of the meson, $G_E(0)=1$ fixing the charge, $G_M(0)=\mu_\rho$
the magnetic moment and $G_Q(0)=\mathcal{Q}_\rho$ the quadrupole moment. The
charge radius comes from the slope of the electric form factor at the origin,
\begin{equation}
r_\rho^2=-6\left.\frac{d}{dQ^2}G_E(Q^2)\right|_{Q^2=0}\,.\label{rcharge}
\end{equation}
Following the conventions of Refs.~\cite{Bhagwat:2006pu, Hedditch:2007ex}, we
quote $\mu_\rho$ in units of $e/(2m_\rho)$ and $\mathcal{Q}_\rho$ in units of
$e/m_\rho^2$, so that a pointlike spin-1 particle would carry $\mu_\rho=2$ and
$\mathcal{Q}_\rho=-1$. Departures from those values measure how far the $\rho$ is
from being structureless, and the sign of $\mathcal{Q}_\rho$ tells whether the
charge distribution is oblate or prolate.

Within the generalized impulse approximation, the vertex is the triangle diagram
\begin{equation}
\Lambda_{\lambda,\mu\nu}(K,Q)=2N_c\,E_\rho^2\,P_T(Q^2)\,
{\rm Tr}\int\frac{d^4q}{(2\pi)^4}\gamma_\nu^T\,S_f(q+p^f)\,
\mi\gamma_\lambda\,S_f(q+p^i)\,\gamma_\mu^T\,S_g(q)\,,\label{LambdaGIA}
\end{equation}
where the amplitude and the transverse dressing have been taken outside the
integral, as their momentum independence permits. Projecting out the invariant
functions and introducing 2 Feynman parameters gives
\begin{equation}
F_i(Q^2)=\frac{3}{4\pi^2}E_\rho^2\int_0^1 d\alpha\int_0^{1-\alpha}d\beta\,
\alpha\left[\mathcal{A}_i\,\overline{\mathcal{C}}_1^{\rm iu}(\omega_2)
+\left(\mathcal{B}_i-\mathcal{A}_i\,\omega_2\right)
\overline{\mathcal{C}}_2^{\rm iu}(\omega_2)\right]\,,\label{FiQ2}
\end{equation}
with
$\omega_2=M_f^2+\alpha(1-\alpha)m_\rho^2+\alpha\beta(1-\alpha-\beta)Q^2$ and
$\overline{\mathcal{C}}_2^{\rm iu}(z)=-d\overline{\mathcal{C}}_1^{\rm iu}(z)/dz$.
The kinematic coefficients read
\begin{eqnarray}
\mathcal{A}_1&=&2-\alpha\,,\\
\mathcal{A}_2&=&\frac{m_\rho^2\left(\alpha(10\beta-7)-4\right)
+Q^2\alpha(2\beta-1)}{2m_\rho^2}\,,\\
\mathcal{A}_3&=&\frac{2\alpha(1-2\beta)(5m_\rho^2+Q^2)}{4m_\rho^2+Q^2}\,,\\
\mathcal{B}_1&=&2\left((\alpha-2)\alpha^2(1-\beta)\beta Q^2
+\alpha M_f^2+2(1-\alpha)M_f M_g\right.\nonumber\\
&&\left.+(1-\alpha)^2\alpha m^2_{\rho}\right)\,,\\
m_\rho^2\,\mathcal{B}_2&=&\alpha Q^2(2\beta-1)M_f M_g
+\alpha^2Q^2m_{\rho}^2\left(2\alpha\beta^3-5\alpha\beta^2
+(\alpha+2)\beta+\alpha-1\right)\nonumber\\
&&+m_{\rho}^2\left(\alpha(6\beta-5)M_f^2
+2M_f M_g(2\alpha\beta-\alpha-2)\right.\nonumber\\
&&\left.-(\alpha-1)\alpha m_{\rho}^2(10\alpha\beta-7\alpha-6\beta+1)\right)\,,\\
(4m_\rho^2+Q^2)\mathcal{B}_3&=&4\alpha m_{\rho}^2\left((3-6\beta)M_f^2
+2(1-2\beta)M_f M_g\right.\nonumber\\
&&\left.+(\alpha-1)m_{\rho}^2\left(16\alpha\beta^2-6(\alpha+1)\beta
-5\alpha+3\right)\right)\nonumber\\
&&-4\alpha Q^2\left((2\beta-1)\left(M_f M_g
+\alpha\beta m_\rho^2(\alpha(\beta-3)+2)\right)\right.\nonumber\\
&&\left.+(\alpha-1)\alpha m_{\rho}^2\right)\,.
\end{eqnarray}
The 2 dressed masses have been kept distinct so that the same expressions can
be applied to unequal-flavor vector states such as the $K^{*}$. For the $\rho$ we
set $M_g=M_f$ throughout.

\section{Numerical results}\label{section-4}

The calculation proceeds in a fixed order. The gap equation, Eq.~(\ref{CI11}),
is solved first for the dressed mass as a function of the flavor number, both in
the chiral limit and with a current mass. That mass enters the kernel
$K_\gamma(Q^2)$ of Eq.~(\ref{Kgamma}), whose pole condition, Eq.~(\ref{rhobse}),
returns $m_\rho$. The canonical normalization of Eq.~(\ref{Erhonorm}) then fixes
$E_\rho$, and from it follow the decay constant of Eq.~(\ref{fpim}) and the form
factors of Eq.~(\ref{FiQ2}). The parameters are those of Table~\ref{tab1}, fitted
in Refs.~\cite{Roberts:2010rn, Chen:2012txa} to the ground-state light mesons at
the physical flavor number. Nothing is refitted when $N_f$ is varied, so every
trend reported below is a consequence of Eq.~(\ref{CI4b}) alone.

\begin{table*}[t]
\caption{ Parameters of the flavor-dependent contact interaction,
used as input to the gap equation and to the Bethe-Salpeter equation. They were
fixed in Ref.~\cite{Roberts:2010rn} by the ground-state properties of the $\pi$
and $\rho$ mesons at $N_f=2$ and $N_c=3$. The trial critical number
$\mathcal{N}^{c}_{f}$ is tuned so that Eq.~(\ref{CI4b}) delivers the desired
critical flavor number \cite{Ahmad:2020jzn}. Dimensionful quantities are in GeV. }
\centering
\begin{tabular}{cccccc}
\hline
$m_{u=d}$ & $\Lambda_{ir}$ & $\Lambda_{uv}$ & $\alpha_{ir}$ & $m_{g}$ &
$\mathcal{N}^{c}_{f}$\\
\hline
$0.007$ & $0.24$ & $0.905$ & $0.93\pi$ & $0.8$ & $10.9$\\
\hline
\hline
\end{tabular}\label{tab1}
\end{table*}

\subsection{Benchmark at the physical flavor number}

Consistency with earlier work has to be established before any extrapolation is
credible. Setting $N_f=2$ and $N_c=3$ we obtain a chiral-limit dressed mass
$M_0=0.358$ GeV and $M_{u/d}=0.368$ GeV with the current mass of
Table~\ref{tab1}. The vector-channel quantities that follow are collected in
Table~\ref{tab2}. They agree with the contact-interaction determination of
Ref.~\cite{Roberts:2011cf} to better than one part in a thousand, which is the
level one expects when the same parameters and the same regularization are used.
The mass overshoots the measured value by roughly twenty percent, a known and
well documented feature of a momentum-independent kernel in the vector channel
rather than a defect of the flavor dependence studied here.

\begin{table*}[t]
\caption{\label{tab2} Ground-state properties of the $\rho$ meson at $N_f=2$ and
$N_c=3$, obtained from Eqs.~(\ref{rhobse}), (\ref{Erhonorm}) and (\ref{fpim}),
compared with the contact-interaction results of Ref.~\cite{Roberts:2011cf} and
with experiment \cite{ParticleDataGroup:2024cfk}. Masses and decay constants are
in GeV and the amplitude $E_\rho$ is dimensionless. }
 \centering
\begin{tabular}{lccc}
\hline
 & $E_{\rho}$ & $f_{\rho}$ & $m_{\rho}$\\
\hline
This work & $1.530$ & $0.129$ & $0.929$\\
Contact interaction \cite{Roberts:2011cf} & $1.520$ & $0.130$ & $0.919$\\
Experiment \cite{ParticleDataGroup:2024cfk} & --- & --- & $0.77526(23)$\\
\hline
\hline
\end{tabular}
\end{table*}

\subsection{Critical flavor number and the loss of confinement}

The chiral limit admits a closed-form answer, and it is worth recording because
it removes any ambiguity about where the transition sits. As $M_f\rightarrow 0$
the regularized integral of Eq.~(\ref{CI12}) approaches a finite constant,
$\mathcal{A}_{01}(M_f^2)\rightarrow\Lambda_{uv}^{2}-\Lambda_{ir}^{2}$, so the
nontrivial solution of Eq.~(\ref{CI11}) disappears exactly when the coupling
drops below
$\alpha_{\rm eff}(N^{c}_{f})=3\pi^{2}/(\Lambda_{uv}^{2}-\Lambda_{ir}^{2})$.
Inverting Eq.~(\ref{CI4b}) gives
\begin{equation}
N^{c}_{f}\Bigg|_{\rm chiral}=2+\mathcal{N}^{c}_{f}
\left[1-\left(\frac{3\pi^{2}}
{\alpha_{\rm eff}\left(\Lambda_{uv}^{2}-\Lambda_{ir}^{2}\right)}\right)^{2}
\right]\,.\label{CI14}
\end{equation}
With the parameters of Table~\ref{tab1} this yields
$\alpha_{\rm eff}(N^{c}_{f})=38.89$ GeV$^{-2}$ and
$N^{c}_{f}\big|_{\rm chiral}=7.89$, consistent with the value quoted in
Ref.~\cite{Ahmad:2020jzn} and with the lower edge of the conformal window
favored by lattice studies. A current mass of $7$ MeV smears the transition, as
it must, and the steepest descent of $M_{u/d}$ then occurs at
$N^{c}_{f}\big|_{m_f\neq 0}=8.05$.

\begin{figure}
\begin{center}
\includegraphics[width=9cm]{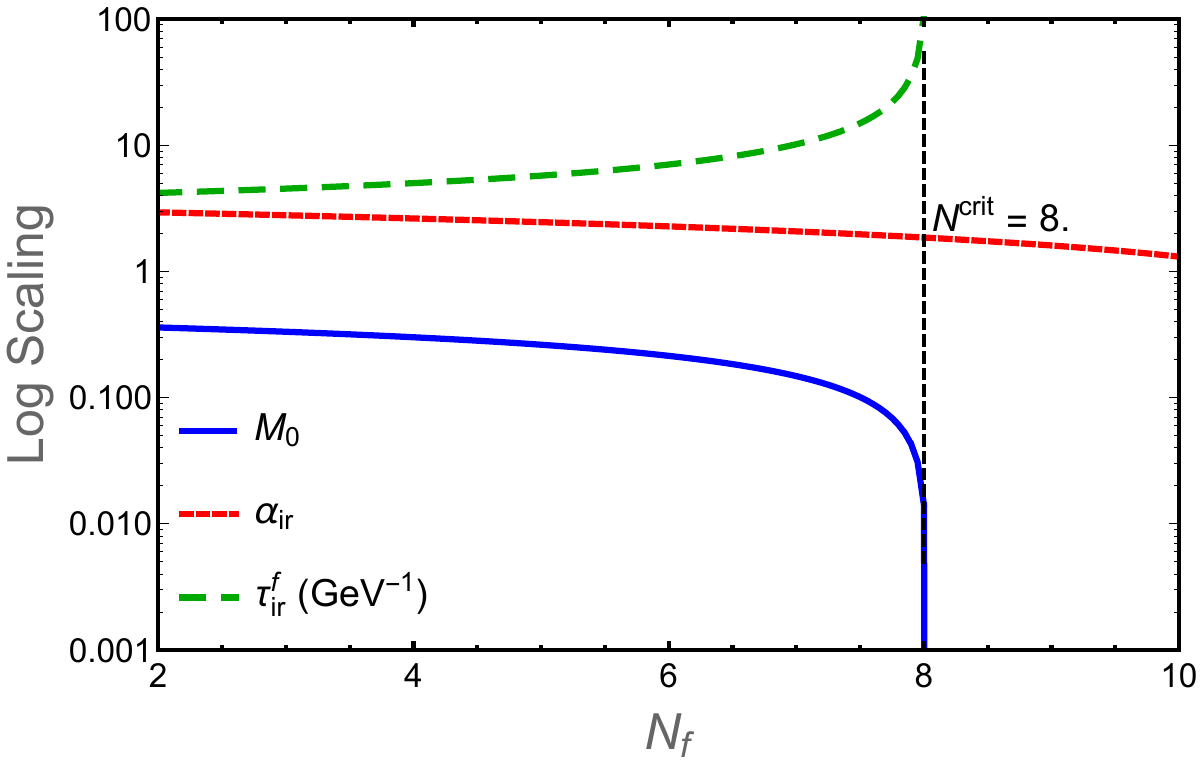}
\caption{Dressed-quark mass and confinement length scale as functions of the
flavor number, in the chiral limit and with $m_{u=d}=0.007$ GeV. The vertical
line marks $N^{c}_{f}\big|_{\rm chiral}=7.89$ from Eq.~(\ref{CI14}), where
$M_0$ vanishes and $\tilde{\tau}_{ir}$ diverges.}
\label{FigMass}
\end{center}
\end{figure}

Confinement follows the same schedule. The length scale of Eq.~(\ref{CI13})
grows slowly at first, reaching $1.62\,\tau_{ir}$ at $N_f=6$ in the chiral limit,
and then runs away, passing $7\,\tau_{ir}$ at $N_f=7.8$ and diverging at
$N^{c}_{f}$. With a current mass the divergence softens into a steep rise, from
$1.52\,\tau_{ir}$ at $N_f=6$ to $12.7\,\tau_{ir}$ at $N_f=10$. Both behaviors are
displayed in Fig.~\ref{FigMass}. Since it is the infrared cutoff that suppresses
the quark production thresholds, its divergence is the statement that those
thresholds have come back.

\subsection{Flavor dependence of the vector bound state}

The behavior of the vector channel is the central result of this work, and it
departs from what the pseudoscalar sector had led one to expect. Table~\ref{tab3}
collects the numbers.

\begin{table*}[t]
\caption{\label{tab3} Flavor dependence of the $\rho$-meson properties, in the
chiral limit and with a current mass. Masses are in GeV and $E_\rho$ is
dimensionless. The confinement length is quoted relative to its value at $N_f=2$.
In the chiral limit no solution survives beyond
$N^{c}_{f}\big|_{\rm chiral}=7.89$.}
\centering
\begin{tabular}{ccccccc}
\hline
$N_{f}$ & $M_{f}$ & $m_{\rho}$ & $m_{\rho}/2M_{f}$ & $E_{\rho}$ & $f_{\rho}$ &
$\tilde{\tau}_{ir}/\tau_{ir}$\\
\hline
\multicolumn{7}{c}{chiral limit, $m_f=0$}\\
\hline
$2$   & $0.3576$ & $0.9184$ & $1.28$ & $1.519$ & $0.130$ & $1.00$\\
$3$   & $0.3321$ & $0.9025$ & $1.36$ & $1.457$ & $0.133$ & $1.08$\\
$4$   & $0.3023$ & $0.8850$ & $1.46$ & $1.392$ & $0.137$ & $1.18$\\
$5$   & $0.2664$ & $0.8657$ & $1.63$ & $1.324$ & $0.141$ & $1.34$\\
$6$   & $0.2207$ & $0.8441$ & $1.91$ & $1.251$ & $0.146$ & $1.62$\\
$7$   & $0.1556$ & $0.8199$ & $2.63$ & $1.173$ & $0.153$ & $2.30$\\
$7.5$ & $0.1047$ & $0.8066$ & $3.85$ & $1.132$ & $0.156$ & $3.42$\\
$7.8$ & $0.0511$ & $0.7982$ & $7.81$ & $1.106$ & $0.159$ & $7.00$\\
\hline
\multicolumn{7}{c}{$m_{u=d}=0.007$ GeV}\\
\hline
$2$  & $0.3675$ & $0.9285$ & $1.26$  & $1.530$ & $0.129$ & $1.00$\\
$3$  & $0.3432$ & $0.9133$ & $1.33$  & $1.468$ & $0.132$ & $1.07$\\
$4$  & $0.3151$ & $0.8966$ & $1.42$  & $1.404$ & $0.136$ & $1.17$\\
$5$  & $0.2820$ & $0.8785$ & $1.56$  & $1.336$ & $0.140$ & $1.30$\\
$6$  & $0.2414$ & $0.8588$ & $1.78$  & $1.264$ & $0.145$ & $1.52$\\
$7$  & $0.1893$ & $0.8383$ & $2.21$  & $1.189$ & $0.151$ & $1.94$\\
$8$  & $0.1199$ & $0.8207$ & $3.42$  & $1.111$ & $0.158$ & $3.07$\\
$9$  & $0.0552$ & $0.8223$ & $7.45$  & $1.040$ & $0.166$ & $6.66$\\
$10$ & $0.0290$ & $0.8440$ & $14.55$ & $0.967$ & $0.174$ & $12.67$\\
\hline
\hline
\end{tabular}
\end{table*}

What stands out first is how little the mass moves. Between $N_f=2$ and
$N_f=10$ the dressed-quark mass falls by a factor of 13, from $0.368$ to
$0.029$ GeV, while $m_\rho$ changes by 12\%. The vector meson
essentially decouples from the constituent scale that produced it. Nothing
protects $m_\rho$, since the $\rho$ is not a Goldstone boson, but nothing drives
it either, and the two effects that might have done so work against each other:
the constituents get lighter while the interaction that binds them gets weaker,
and in this model the cancellation is nearly complete. The mass reaches a shallow
minimum of $0.818$ GeV at $N_f=8.44$, just above the critical flavor number, and
climbs again beyond it. That upturn is real but modest, and it would be a mistake
to read it as the dramatic rise seen in the pseudoscalar channel, where the
Goldstone mass is tied directly to the order parameter of the symmetry being
restored.

\begin{figure}
\begin{center}
\includegraphics[width=9cm]{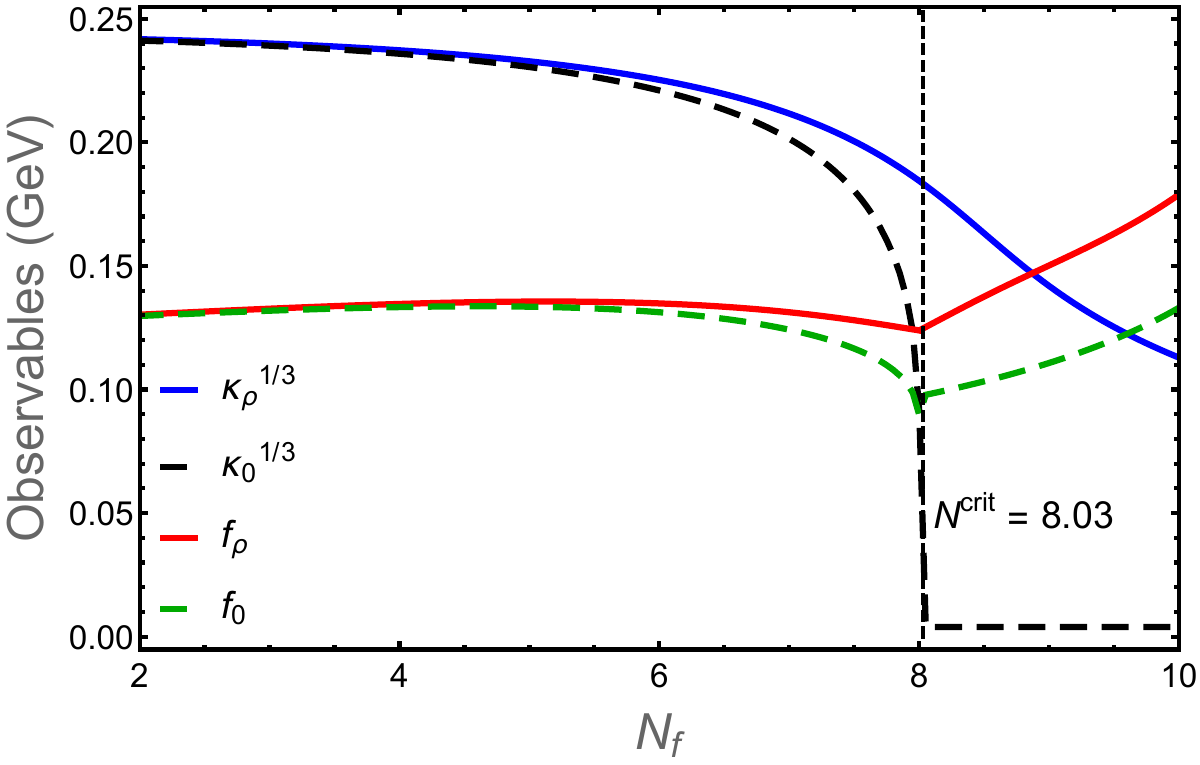}
\includegraphics[width=9cm]{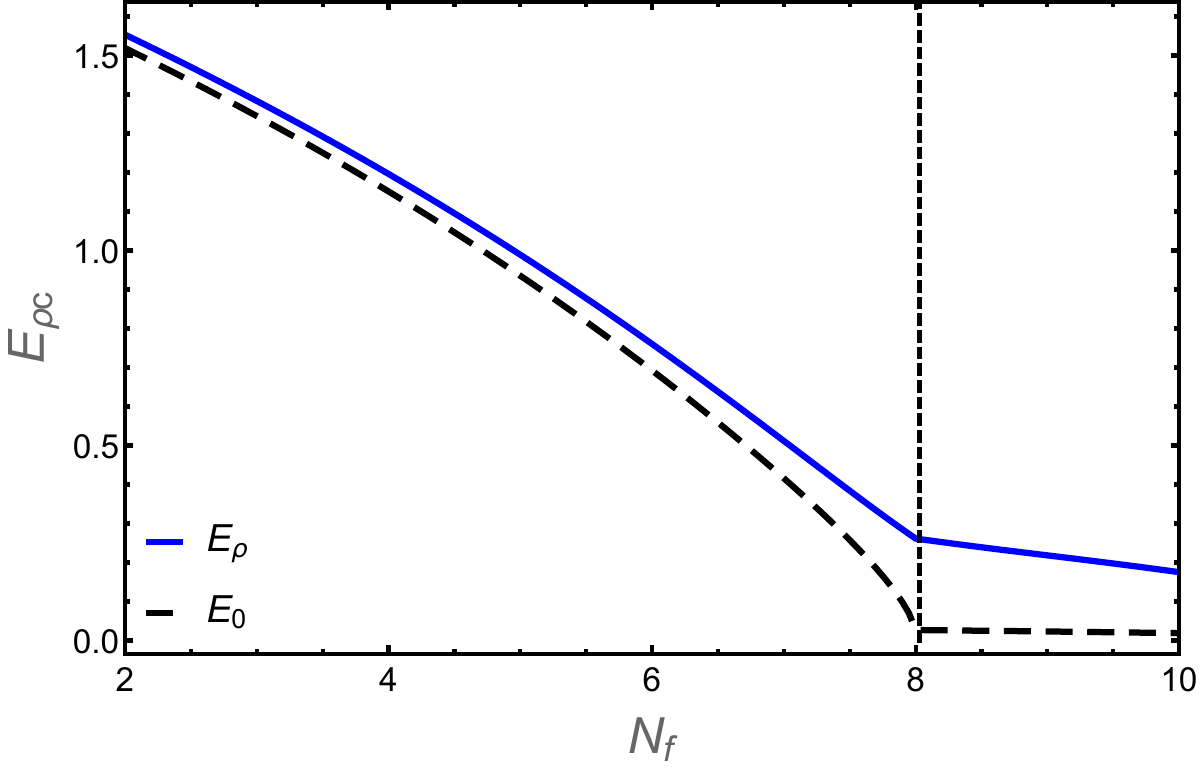}
\caption{Mass $m_\rho$, leptonic decay constant $f_\rho$ and canonically
normalized amplitude $E_\rho$ as functions of the flavor number, in the chiral
limit and with a current quark mass.}
\label{FigFrho}
\end{center}
\end{figure}

The amplitude and the decay constant separate cleanly, as Fig.~\ref{FigFrho}
shows. The canonically normalized $E_\rho$ falls by 37\% over the same interval, from
$1.530$ to $0.967$, which is what one expects from a state that becomes
progressively less tightly bound. The decay constant moves the other way,
drifting upward by about 1/3, from $0.129$ to $0.174$ GeV. The 2 are not in
conflict. Equation~(\ref{fpim}) makes $f_\rho$ proportional to
$E_\rho/(m_\rho\alpha_{\rm eff}(N_f))$, and the coupling in the denominator is
screened away faster than the amplitude in the numerator shrinks. This is the
single most model-dependent statement in the present analysis, because it hangs
entirely on the explicit inverse coupling in the residue of the vector pole, and
it deserves to be tested against a momentum-dependent kernel before being taken
as a property of the theory rather than of the interaction.

\subsection{Two criteria for dissociation}

At finite temperature a hadron is said to melt at the Mott point, where its mass
crosses the sum of the masses of its constituents and the decay into them opens
up \cite{Hufner:1996pq, Contrera:2009hk, Xu:2021lxa}. Transplanting that
criterion to the flavor number is the natural thing to try, and the fourth column
of Table~\ref{tab3} shows why it cannot be done here. The ratio $m_\rho/2M_f$
never approaches unity from below. It starts at $1.26$ at the physical flavor
number, grows through $1.78$ at $N_f=6$, reaches $3.42$ at $N_f=8$ and $14.55$ at
$N_f=10$. The $\rho$ of the contact interaction sits above its nominal
quark-antiquark threshold everywhere, including at the point where the model was
fitted, and the gap widens monotonically. A criterion built on a crossing of
$m_\rho$ and $2M_f$ therefore never triggers, and the growth of the ratio is
driven almost entirely by the collapse of $M_f$ rather than by any motion of the
bound state.

That the state nevertheless exists is not a contradiction. The infrared cutoff
of the proper-time scheme removes the production threshold altogether, so there
is no continuum for the pole to dissolve into, and the excess energy has nowhere
to go. Confinement, in this model, is what keeps a state that lies above its own
threshold from decaying. The quantity that decides the fate of the $\rho$ is
therefore not the mass but the confinement length of Eq.~(\ref{CI13}). When
$\tilde{\tau}_{ir}$ diverges at $N^{c}_{f}$, the thresholds reopen, and the
excess that was present all along becomes an open decay channel in a single
stroke. The dissociation flavor number coincides with the critical one,
$N^{d}_{\rho f}=N^{c}_{f}$, but for a reason that has nothing to do with a
crossing of curves.

The comparison between the two criteria is worth stating plainly, because it
sharpens what the model can and cannot say. In the pseudoscalar channel both
criteria agree, since the Goldstone mass rises steeply through the threshold
just as confinement is lost \cite{Ahmad:2024emu}. In the vector channel they come
apart, one of them never fires, and only the confinement criterion retains
meaning. Whether that separation survives a momentum-dependent kernel, which
would place the $\rho$ below $2M_f$ to begin with, is an open question and one of
the more interesting consequences of the present calculation.

\subsection{Electromagnetic form factors}

We evaluate $G_E$, $G_M$ and $G_Q$ from Eqs.~(\ref{GEdef}) to (\ref{GQdef}) and
Eq.~(\ref{FiQ2}) for flavor numbers below the critical one. Above $N^{c}_{f}$
the generalized impulse approximation of Eq.~(\ref{LambdaGIA}) presupposes a
stable external state with a canonically normalized amplitude, a condition that
no longer holds once the thresholds have reopened, so the form-factor analysis
stops at $N_f=8$ and the region beyond it is characterized only through the mass
trajectory. The results are normalized to unit electric charge, which is the
standard convention for the static moments and has the practical advantage that
the overall constant of Eq.~(\ref{FiQ2}) drops out of every quantity reported in
Table~\ref{tab4}.

\begin{table*}[t]
\caption{\label{tab4} Static electromagnetic properties of the $\rho$ meson as
functions of the flavor number, with $m_{u=d}=0.007$ GeV. The magnetic moment is
in units of $e/(2m_\rho)$ and the quadrupole moment in units of $e/m_\rho^2$, so
that a pointlike spin-1 particle would give $\mu_\rho=2$ and
$\mathcal{Q}_\rho=-1$. The charge radius is in fm and the zero crossing of the
electric form factor in GeV$^2$.}
\centering
\begin{tabular}{ccccc}
\hline
$N_{f}$ & $\mu_{\rho}$ & $\mathcal{Q}_{\rho}$ & $r_{\rho}$ &
$Q^{2}_{\rm node}$\\
\hline
$2$ & $2.291$ & $-0.695$ & $0.334$ & $3.539$\\
$3$ & $2.295$ & $-0.695$ & $0.341$ & $3.379$\\
$4$ & $2.300$ & $-0.696$ & $0.347$ & $3.208$\\
$5$ & $2.305$ & $-0.697$ & $0.355$ & $3.024$\\
$6$ & $2.312$ & $-0.699$ & $0.365$ & $2.827$\\
$7$ & $2.320$ & $-0.700$ & $0.375$ & $2.621$\\
$8$ & $2.331$ & $-0.702$ & $0.386$ & $2.434$\\
\hline
\hline
\end{tabular}
\end{table*}

\begin{figure}
\begin{center}
\centering
\begin{subfigure}[b]{\textwidth}
\centering
\includegraphics[width=9cm]{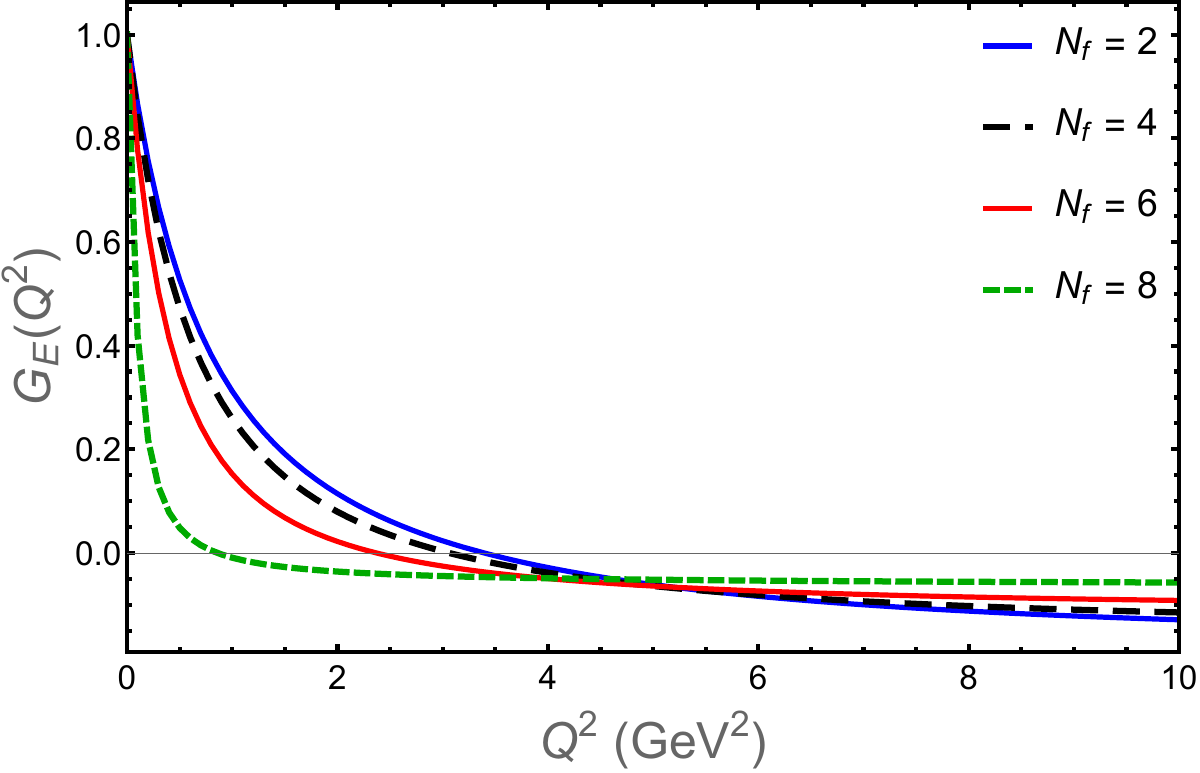}
\caption{}
\label{fig:ge}
\end{subfigure}

\vspace{0.5cm}

\begin{subfigure}[b]{\textwidth}
\centering
\includegraphics[width=9cm]{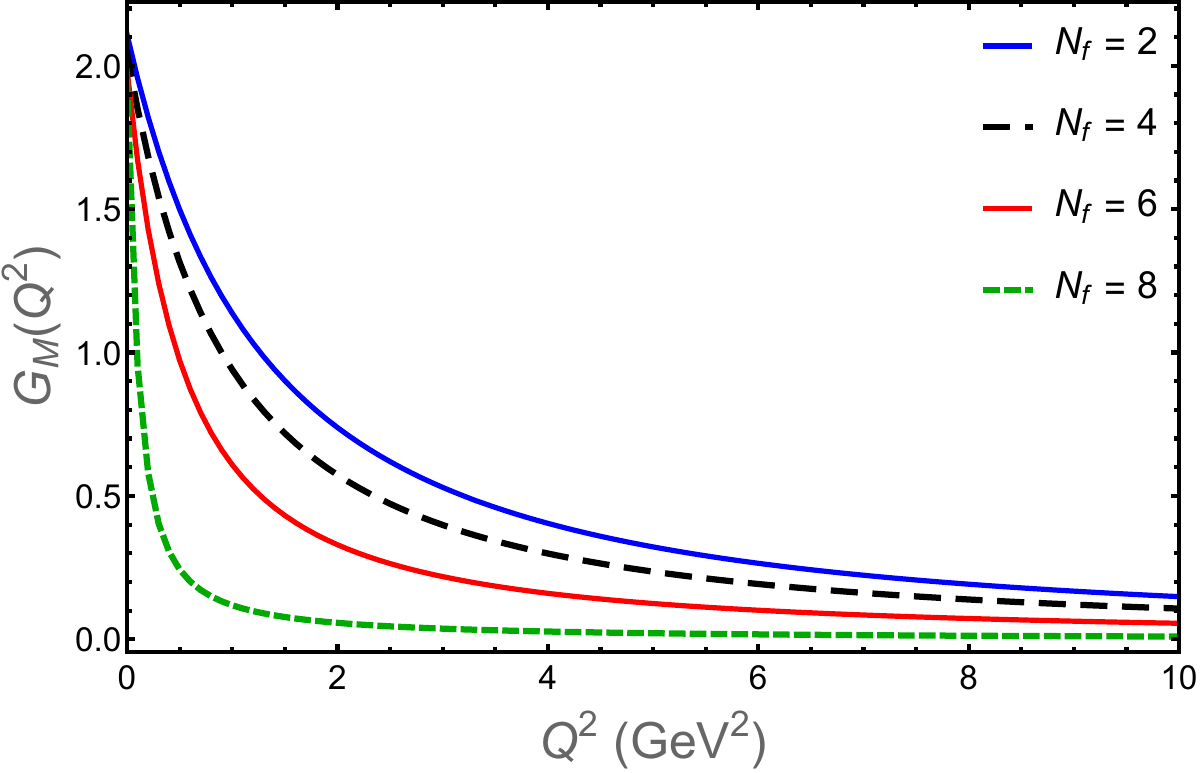}
\caption{}
\label{fig:gm}
\end{subfigure}

\vspace{0.5cm}

\begin{subfigure}[b]{\textwidth}
\centering
\includegraphics[width=9cm]{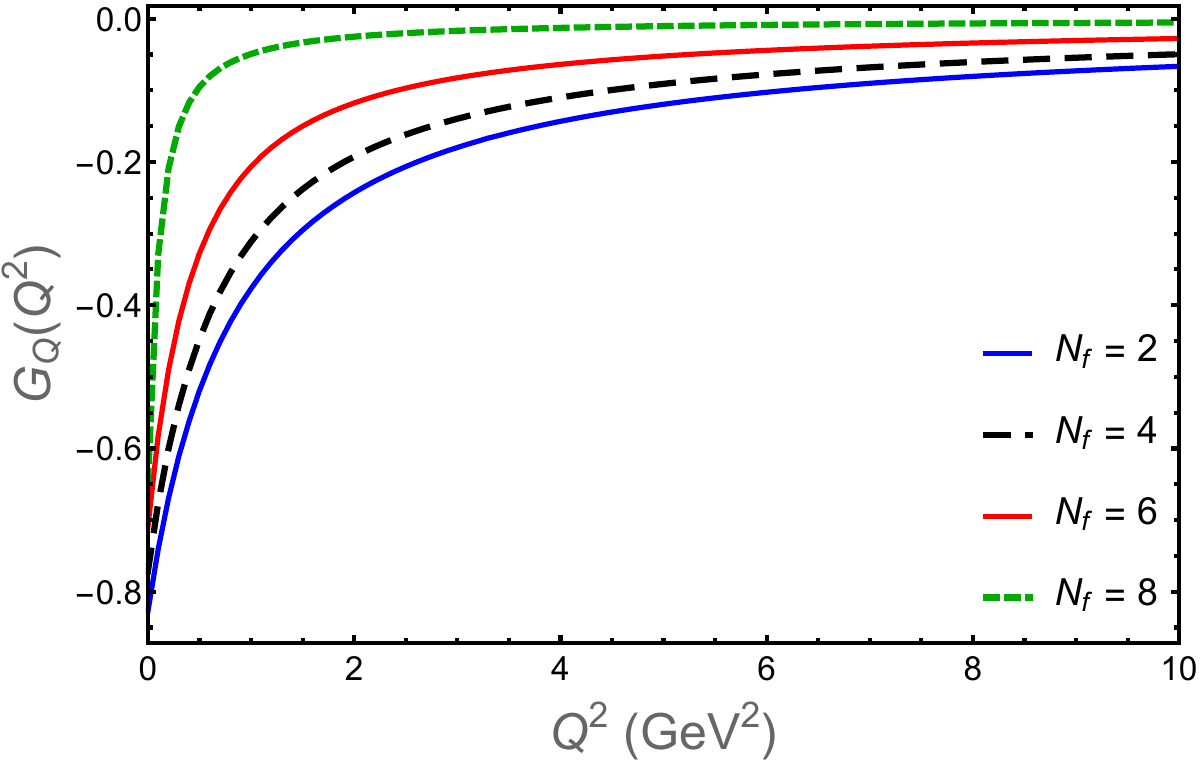}
\caption{}
\label{fig:gq}
\end{subfigure}
\caption{Elastic electromagnetic form factors of the $\rho$ meson, normalized to
unit charge, for $N_f=2$, $4$, $6$ and $8$. Panel (a) shows $G_E(Q^2)$, panel (b)
$G_M(Q^2)$ and panel (c) $G_Q(Q^2)$. The lattice determinations of
Refs.~\cite{Hedditch:2007ex, Owen:2015gva} and the Schwinger-Dyson calculation of
Ref.~\cite{Bhagwat:2006pu} are included for comparison at $N_f=2$.
}

\label{FigFF}
\end{center}
\end{figure}

\begin{figure}
\begin{center}
\includegraphics[width=9cm]{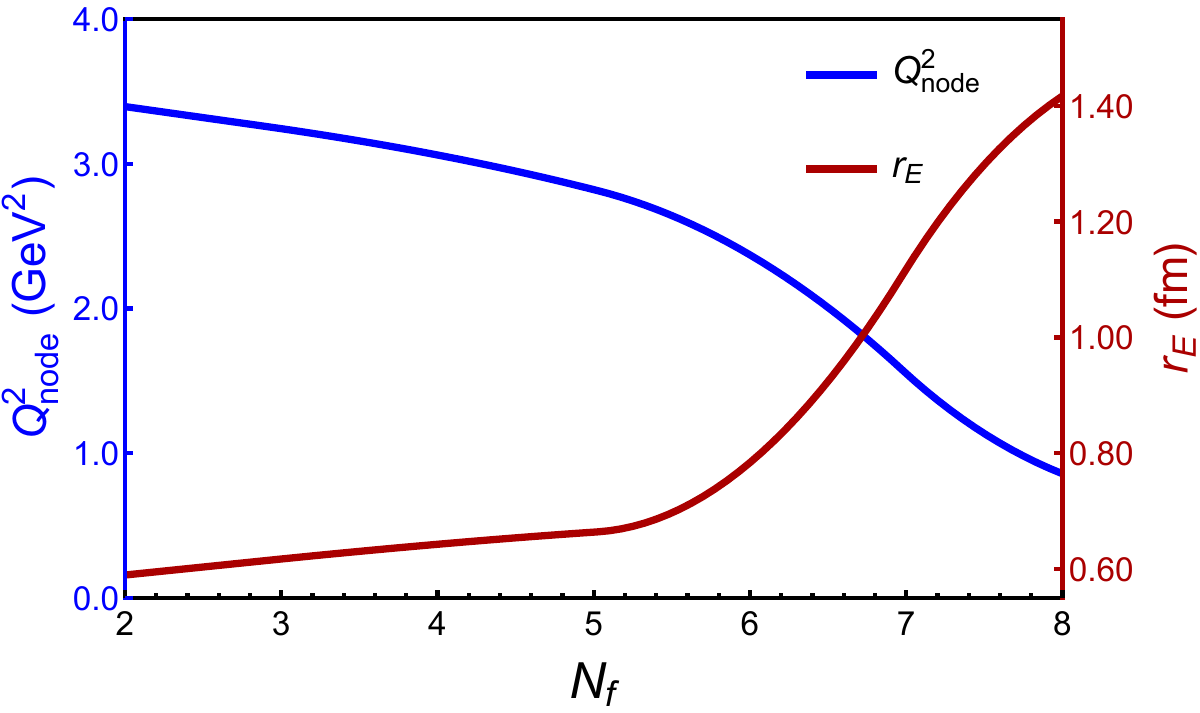}
\caption{Charge radius, magnetic moment and quadrupole moment of the $\rho$
meson as functions of the flavor number.
}
\label{FigStatic}
\end{center}
\end{figure}

The mechanism behind these trends is visible in the integrand. The scale that
controls Eq.~(\ref{FiQ2}) is
$\omega_2=M_f^2-\alpha(1-\alpha)m_\rho^2+\alpha\beta(1-\alpha-\beta)Q^2$, and
since $M_f$ drops steadily with the flavor number while $m_\rho$ barely moves,
the momentum window over which the form factors retain strength narrows. All
3 soften, the charge radius expands by 16\% between $N_f=2$ and
$N_f=8$, and the zero crossing of the electric form factor migrates towards the
infrared by nearly 1/3, from $3.54$ to $2.43$ GeV$^2$. Figure~\ref{FigFF}
shows the 3 form factors and Fig.~\ref{FigStatic} the static moments that
follow from them. A node that moves inward and a radius that grows are two
readings of the same fact, that a meson built from lighter and more loosely
bound constituents occupies more space.

The static moments deserve a separate remark, because they move far less than
the radius does. Both $\mu_\rho$ and $\mathcal{Q}_\rho$ shift by under 2\% across 
the whole range, staying near $2.3$ and $-0.70$ respectively. The
magnetic moment sits above the pointlike value of $2$ and the quadrupole moment
is negative, so the charge distribution is oblate, in agreement with the quenched
lattice determination of Ref.~\cite{Hedditch:2007ex}, with the near-physical-mass
calculation of Ref.~\cite{Owen:2015gva} and with the Dyson-Schwinger results of
Ref.~\cite{Bhagwat:2006pu}. The confining Nambu-Jona-Lasinio treatment of
Ref.~\cite{Carrillo-Serrano:2015uca} reported tension with the lattice precisely
in the quadrupole sector, and our finding that $\mathcal{Q}_\rho$ is essentially
flat in $N_f$ suggests that the discrepancy is a feature of the interaction
rather than of the flavor content.

\begin{figure}
\begin{center}
\includegraphics[width=9cm]{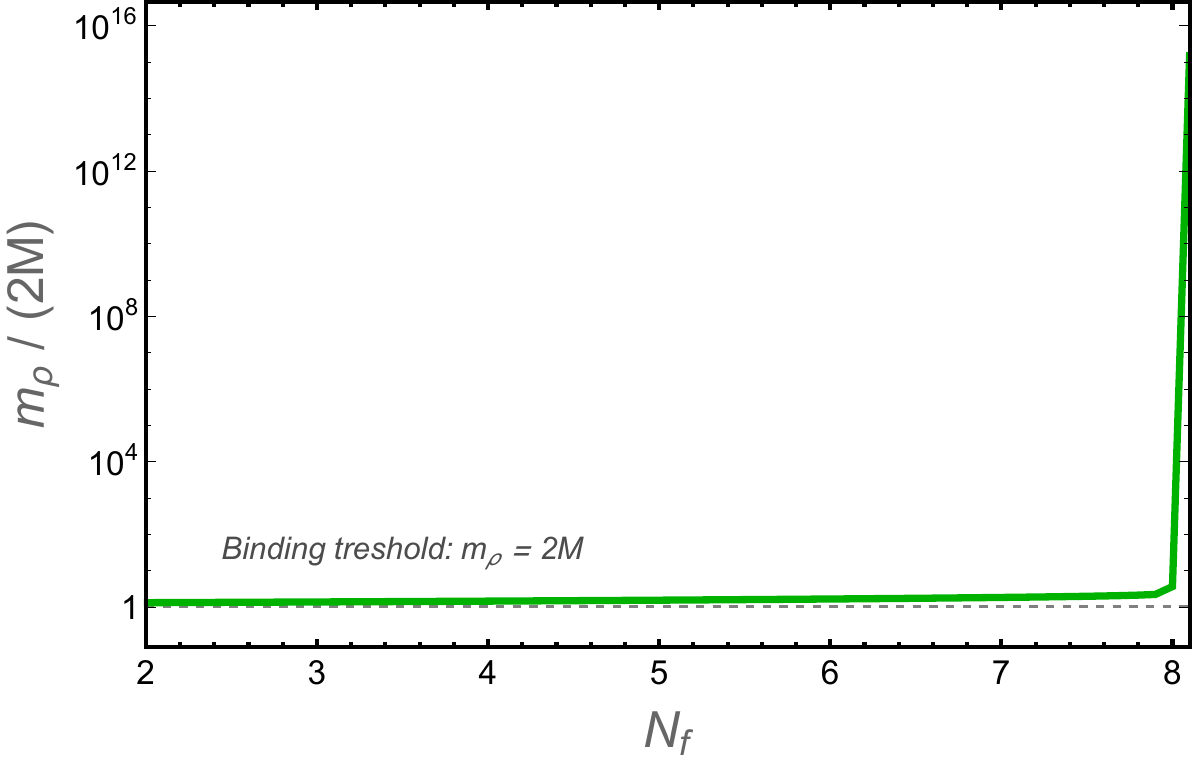}
\caption{\label{fig:mass_ratio} Evolution of the ratio $m_\rho / (2M)$ as a function of the number of flavors $N_f$ shown on a logarithmic scale. The horizontal dashed line denotes the non-relativistic two-body constituent threshold, $m_\rho / (2M) = 1$. The strictly monotonic growth highlights the disparate rates of descent between the dynamically generated dressed-quark mass $M$ and the vector-meson mass $m_\rho$ as the critical flavor threshold $N_f^c$ is approached.}
\end{center}
\end{figure}

Figure~\ref{fig:mass_ratio} depicts the ratio of the $\rho$-meson mass to twice the dynamically generated constituent quark mass, $m_\rho / (2M)$, across the range $2 \le N_f \le 8$. Throughout the investigated domain, the ratio remains strictly greater than unity and exhibits a monotonic increase with $N_f$. For $N_f = 2$, the meson lies just above the non-relativistic two-quark continuum threshold ($m_\rho \gtrsim 2M$), reflecting the resonant nature of the $\rho$-meson within the symmetry-preserving contact interaction framework. As $N_f$ increases, quark-flavor screening progressively diminishes the infrared coupling strength $\alpha_{\text{ir}}(N_f)$. Because dynamical chiral symmetry breaking is inherently non-linear, the dressed-quark mass $M$ drops precipitously toward the bare current-quark mass $m_c$. Conversely, the vector-meson bound-state mass $m_\rho$, constrained by the Bethe-Salpeter eigenvalue equation, exhibits considerable dynamical inertia and decreases at a substantially slower rate. Consequently, the rapid shrinkage of the denominator triggers an exponential-like escalation in $m_\rho / (2M)$ as $N_f \to N_f^c$, serving as a distinctive structural signature of the impending restoration of chiral symmetry.

\section{Summary and perspectives}\label{section-5}

We have followed the ground-state properties of the $\rho$ meson as the number
of light-quark flavors is driven towards the conformal window, using the
flavor-dependent contact interaction as input to the Schwinger-Dyson equation for
the dressed quark and to the homogeneous Bethe-Salpeter equation in rainbow-ladder
truncation, regularized in the Schwinger proper-time scheme. Screening from
fermion loops enters through a single square-root dependence of the effective
coupling on the flavor number, and no parameter is refitted along the way.

The chiral limit admits the closed expression of Eq.~(\ref{CI14}) for the
critical flavor number, which places it at $7.89$ for the standard parameter set
and reproduces the value obtained previously by other means. Away from that
limit, a current mass of $7$~MeV moves the steepest descent of the dressed mass
to $8.05$.

The vector channel turned out to behave in a way the pseudoscalar results did not
anticipate. The mass of the $\rho$ is remarkably insensitive to the flavor
number, changing by 12\% while the constituent mass falls by a factor
of 13, with a shallow minimum just above the critical point. The
canonically normalized amplitude decreases steadily and the leptonic decay
constant drifts upward, a separation traceable to the explicit inverse coupling
in the residue of the vector pole. The elastic form factors soften as the
constituent scale shrinks, the charge radius grows and the zero crossing of the
electric form factor migrates towards the infrared, while the magnetic and
quadrupole moments stay essentially where they were.

The most consequential finding is about the criterion for dissociation itself. In
this framework the $\rho$ lies above the sum of the constituent masses at every
flavor number examined, including the physical one, so a Mott-like criterion
based on the crossing of $m_\rho$ and $2M_f$ never applies. What holds the state
together is the removal of the production thresholds by the infrared regulator,
and what destroys it is the divergence of the confinement length at $N^{c}_{f}$.
The two criteria, which coincide in the pseudoscalar channel, come apart here,
and only the second one carries information.

Two limitations bound these conclusions. A momentum-independent amplitude makes
the form factors harder at large momentum transfer than a momentum-dependent
kernel would, so the flavor trends are more trustworthy than the absolute values,
and a kernel with realistic momentum dependence would place the $\rho$ below
$2M_f$ from the start, which may restore the crossing criterion and change the
picture we have just described. The flavor dependence of the coupling is also a
modelling device for the infrared, tuned to the behavior of the dynamical mass
rather than derived, and confronting it with lattice studies of many-flavor gauge
theories is the natural test.

Several continuations suggest themselves. The $K^{*}$ would activate the tensor
covariant that vanishes here, since unequal constituent masses remove the
degeneracy that forces $F_\rho$ to zero, and the expressions of
Sec.~\ref{section-3} were kept general enough to accommodate it. The transition
form factors connecting the vector and pseudoscalar sectors would probe the same
screening dynamics in a channel where both behaviors we have described meet.
Going beyond rainbow-ladder truncation would settle how much of the flavor
dependence reported here survives once $F_\rho$ is no longer forced to vanish.

\bmhead{Acknowledgements}

We thank A. Bashir and B. Masud for guidance and for many useful suggestions, and
our colleagues at the Institute of Physics, Gomal University, for their
hospitality and support.

\section*{Declarations}

\begin{itemize}
\item \textbf{Funding}: AR acknowledges support from SECIHTI (Mexico) under grant CBF-2026-465 
and from UMSNH-CIC under grant 18371.
\item \textbf{Conflict of interest}: the authors declare no competing interests.
\item \textbf{Data availability}: this is a theoretical study and no experimental data
were used. The numerical values reported in Tables~\ref{tab2} to \ref{tab4}, and
the code that generates them, are available from the corresponding author on
reasonable request.
\item \textbf{Author contributions}: \\
Aftab Ahmad designed the study, supervised the research and wrote the original draft. \\ Samreen Fatima and Muhammad Shahbaz contributed in analytical calculation, tabulating and  prepared  some plots. \\ Marco Antonio Bedolla contributed in analytical calculations, prepared some figures and  revised the draft. \\
Alfredo Raya contributed in supervision, write up, revision and provided the funding support.\\ Muhammad Imran revised the manuscript.\\
All authors reviewed and edited the manuscript. 
\end{itemize}

\bibliography{sn-bibliography}

\end{document}